\documentclass[a4paper,fleqn,usenatbib]{mnras}

\usepackage{newtxtext,newtxmath}

\usepackage[T1]{fontenc}
\usepackage{ae,aecompl}

\usepackage{graphicx}	
\usepackage{amsmath}	
\usepackage{amssymb}	

\title[]{Gas kinematics of key prebiotic molecules in GV Tau N  revealed with an ALMA, PdBI, and Herschel synergy}

\author[A. Fuente et al.]{
A. Fuente,$^{1}$\thanks{E-mail: a.fuente@oan.es}
S. P. Trevi\~no-Morales$^{2}$,
R. Le Gal$^3$, 
P. Rivi\`ere-Marichalar$^{1}$,
P. Pilleri$^{4}$,
\newauthor
M. Rodr\'{\i}guez-Baras$^{1}$,
and D. Navarro-Almaida$^1$
\\
$^{1}$Observatorio Astron\'omico Nacional (OAN,IGN), Alfonso XII, 3,  28014, Madrid, Spain\\
$^{2}$Chalmers University of Technology, Department of Space, Earth and Environment, SE-412 93 Gothenburg, Sweden\\
$^{3}$Center for Astrophysics \textbar Harvard \& Smithsonian, 60 Garden St., Cambridge, MA 02138, USA\\
$^{4}$Institut de Recherche en Astrophysique et Planétologie, 9 avenue du colonel Roche, 31028 Toulouse Cedex 4, France\\
}

\date{Accepted XXX. Received YYY; in original form ZZZ}

\pubyear{2020}

\begin{document}
\label{firstpage}
\pagerange{\pageref{firstpage}--\pageref{lastpage}}
\maketitle

\begin{abstract}
A large effort has been made to detect warm gas in the planet formation zone of circumstellar discs
using space and ground-based near infrared facilities. GV Tau N, the most obscured component of the GV Tau system,
is an outstanding source, being one of the first targets detected in HCN and the only one detected in CH$_4$ so far. Although near infrared
observations have shed light on its chemical content, the physical structure and kinematics of the circumstellar matter remained
unknown. We use interferometric images of the HCN 3$\rightarrow$2 and  $^{13}$CO 3$\rightarrow$2 lines, and  far-IR observations
of $^{13}$CO, HCN, CN and H$_2$O transitions to discern the morphology, kinematics, and chemistry of the dense gas close
to the star. These observations constitute the first detection of H$_2$O towards GV Tau N. Moreover, 
ALMA high spatial resolution ($\sim$ 7 au) images of the continuum at 1.1 mm and 
the HCN 3$\rightarrow$2 line resolve different gas components towards GV Tau N, a gaseous disc
with R$\sim$25 au, an ionized jet, and one (or two) molecular outflows. The asymmetric morphology of the gaseous disc shows 
that it has been eroded by the jet. All observations can be explained if GV~Tau~N is binary, and the primary component has a highly inclined individual disc
relative to the circumbinary disc. We discuss the origin of the water  and the other molecules 
emission according to this scenario. In particular, we propose that the water emission would come from the disrupted gaseous disc and the molecular outflows.
\end{abstract}




\section{Introduction}
\label{sec:intro}

A fraction of the gas and dust in protoplanetary discs will end up in planets and may constitute
the basis to form prebiotic species. A large effort has been made to detect the warm gas in the planet formation zone
using Spitzer and ground-based facilities such as the Very Large Telescope (VLT) and the Keck 
Observatory telescopes. In a pioneering work, 
\citet{Lahuis2006} detected strong HCN absorption toward one source, IRS~46, from a sample of more than
100 Class I and II sources located in nearby star-forming regions. Later, \citet{Gibb2007} detected the HCN absorption in GV Tau.
Since then, the observation of absorption and emission lines have provided a view of the molecular  content of the 
very inner regions of protoplanetary discs with several detections of  molecules such as CO, CO$_2$, H$_2$O, OH, HCN, and C$_2$H$_2$ 
\citep{Salyk2007, Salyk2008, Salyk2011, Salyk2019, Carr2008, Carr2011,Carr2014, Pontoppidan2010, Najita2010, Najita2013,
Kruger2011, Doppmann2008, Doppmann2011, Fedele2011, Mandell2012, Bast2013, Gibb2013, Sargent2014, Banzatti2017}.
In addition, the organic species CH$_4$ has been detected towards GV Tau N \citep{Gibb2013}.
In many cases the profiles of the detected lines are broad and centrally peaked \citep{Mandell2012, Salyk2019}. 
This kind of centrally peaked profile is observed even for discs with previously determined high inclination angle and has been interpreted
as emission from a disc wind. By now, direct imaging of the circumstellar material remains difficult with current telescopes, which challenges the interpretation.

The launch of the Herschel Space Observatory allowed to investigate the chemical content of discs at far-IR wavelengths \citep{Thi2011, Salinas2016}, 
providing valuable information on the abundance of H$_2$O in large disc samples
\citep{Hogerheijde2011,Riviere2012, Meeus2012, Fedele2012, Fedele2013, Podio2013, Salinas2016, Alonso2017}. 
Warm water coming from the inner R$\sim$2$-$3 au disc region was detected through the 63.3 $\mu$m line observed with PACS \citep{Riviere2012, Meeus2012}.
The emission of the ground ortho and para water lines are thought to originate in the disc surfaces where UV radiation desorbs the water from the
icy mantles \citep{Hogerheijde2011, Podio2013, Salinas2016}.  In Class I sources, a fraction of the warm water can be emitted in shocks along the outflow cavity as well
\citep{Kristensen2012, Karska2013, Mottram2014}. Thus, the study of far-IR H$_2$O lines provides essential information to characterize the H$_2$O abundance 
from the inner to the cold outer regions of discs.  Moreover, several guaranteed and open time Herschel Key Programs:  
"Water In star-forming regions with Herschel" (WISH; \citealp{vanDishoeck2011}),
"CHEmical Survey of Star-forming regions" (CHESS; \citealp{Ceccarelli2010}),
"Dust, Ice and Gas in Time" (DIGIT; \citealp{Green2013}),
and "GAS in protoplanetary systems (GASPS  \citealp{Mathews2010, Dent2013}), 
have provided information on water in a large number of low-mass protostars
at different evolutionary stages, allowing us to figure out how water is transported from the 
molecular cloud to planet-forming discs.

\begin{figure*}
       \includegraphics [width=0.95\textwidth] {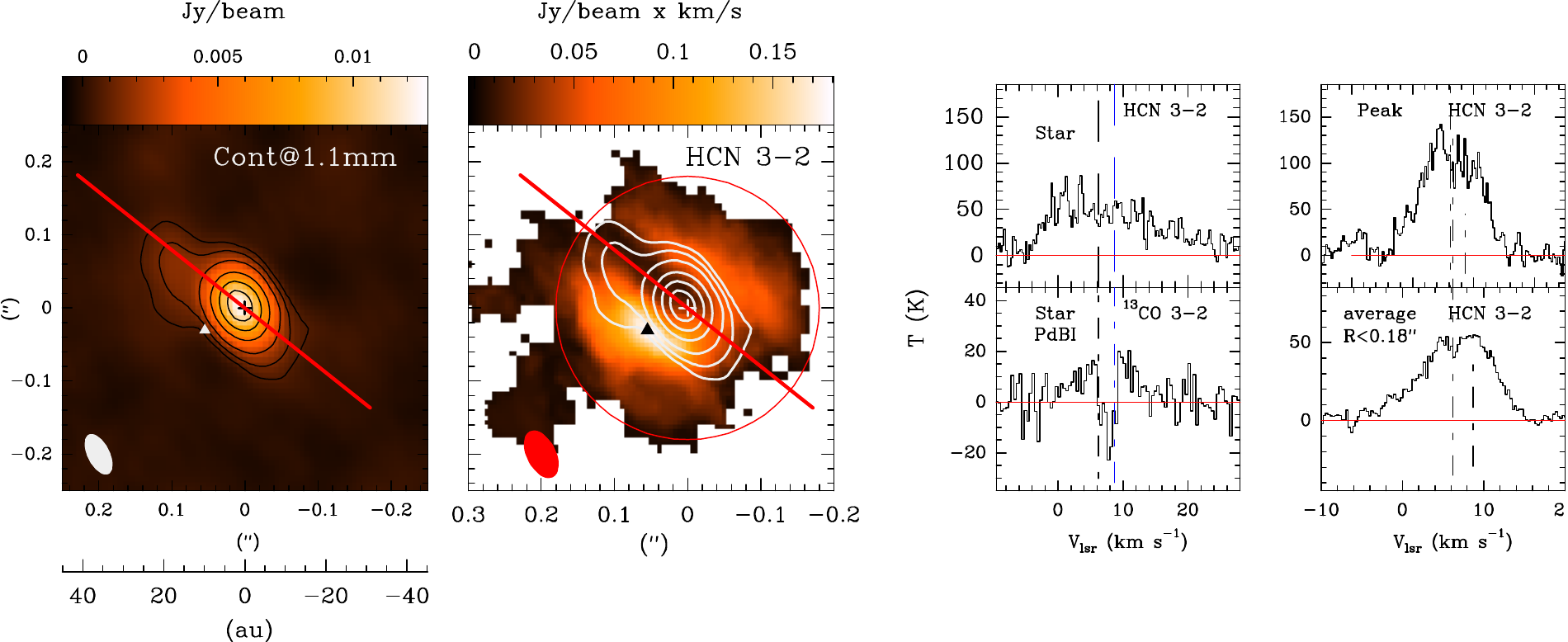}
    \caption{{\it Left:} Continuum image at 1.1~mm as observed with ALMA. Levels are 5$\sigma$, 10$\sigma$, 20$\sigma$ to 100$\sigma$ by 20$\sigma$ where
    $\sigma$=0.13 mJy/beam. 
    {\it Center:}Zero moment map of the HCN 3$\rightarrow$2 line in units of Jy/beam$\times$km~s$^{-1}$. 
    Contour levels are 0.01, 0.025, 0.05, 0.1, 0.15, and 0.175 Jy/beam$\times$km/s. The red line indicates the jet direction derived from the 1.1mm continuum observations, and the circle marks the radius of the gaseous disc, R=0.18$"$. In white, the contours of the 1.1 mm continuum emission.
      {\it Right:} Interferometric spectra of the HCN 3$\rightarrow$2 and $^{13}$CO 3$\rightarrow$2 lines towards the star position (black and white crosses in left and center panels, respectively) as observed with ALMA and PdBI. The HCN 3$\rightarrow$2 spectrum at the emission peak in the gaseous disc (white and black triangles
      in the left and center panels) and 
      the averaged HCN 3$\rightarrow$2 spectrum in the region R$<$0.18$"$ are shown in the right panel. The intensity scale in the spectra is main beam brightness temperature. Vertical lines mark v$_{\rm lsr}$=6.3~km~s$^{-1}$ and 8.7~km~s$^{-1}$.}
    \label{Fig1}
\end{figure*}

The chemical composition of discs in Class I protostars is receiving increasing attention.
The presence of discs in all Class 0 protostars (ages$\sim$0.2~Myr)
is still debated (see e.g.  \citealp{Maret2020}). If indeed discs are present, their masses have been suggested to be high 
($\sim$0.05$-$0.1~M$_\odot$; \citealp{Jorgensen2009}) and they might be gravitationally unstable. Conversely,
the masses of Class II discs in Taurus and Orion (ages$\sim$1$-$5 Myr) have been well studied and are
found to have a median mass of 0.001 M$_\odot$ (e.g. \citealp{Eisner2008, Andrews2013,Tobin2020}). 
These median masses are low compared with the amount of matter needed to form giant
planets, estimated to be 0.01$-$0.1 M$_\odot$ (e.g. \citealp{Weidenschilling1977, Desch2007}).
Class I discs would then be in a transitional stage between massive, highly unstable 
protoplanetary discs to stable discs in which planet formation is progressing. The discs around Class I
objects may therefore accurately represent the initial mass budget for forming planets.

GV~Tau (Haro 6-10) is a T Tauri binary system embedded in the L1524 molecular cloud (d=180$\pm$17~pc) \citep{GAIA2018}.
It is one of the few young binary systems with the primary source (GV~Tau~S) optically visible, and the companion (GV~Tau~N), 
located 1.2$"$ to the North, obscured. It is associated with a parsec-scale Herbig-Haro flow which extends for
1.6~pc to the North and $\sim$1~pc to the South 
\citep{Devine1999, Movsessian1999}. Both stars have been classified as Class I objects with
masses of M$\sim$0.8 M$_\odot$ and ages of $\sim$ 3 Myr \citep{Doppmann2008}. 
Only the GV~Tau~N disc has been detected in the HCN and C$_2$H$_2$ absorption ro-vibrational lines \citep{Gibb2007, Gibb2008}. 
Indeed GV Tau N is an outstanding source, being one of the first targets detected in HCN and the only one detected in CH$_4$  \citep{Gibb2013}.
Moreover, the mass of its protoplanetary disc falls near the
Minimum Mass Solar Nebula, and it may have just enough mass to form giant planets \citep{Sheehan2014}.
Being a Class I disc with a chemistry rich in organic species, GV~Tau~N is a priority target to investigate how organics are
delivered from the gas rich disc to planets. \citet{Fuente2012} reported interferometric images of 
the HCN 3$\rightarrow$2  and HCO$^+$ 3$\rightarrow$2 lines at an angular resolution 
of  $\sim$50~au. Only the HCN 3$\rightarrow$2 emission was detected towards GV Tau N, and was interpreted 
as coming from the unresolved planet forming region.

This paper explores the physical structure, kinematics, and chemistry of the gas in the planet forming region of  GV~Tau~N based on
high spatial resolution images of the HCN 3$\rightarrow$2 and the $^{13}$CO 3$\rightarrow$2 lines, and 
far-IR observations of  the high excitation lines of $^{13}$CO, H$_2$O, HCN and CN.
In particular, we present the first water detection in this system.

\section{Observations} 
\label{sec:obs}

GV Tau was observed with the IRAM\footnote{IRAM is supported by INSU/CNRS (France), MPG (Germany), and IGN (Spain)} Plateau de Bure Interferometer (PdBI), the Atacama Large Millimeter/sub-millimeter  Array (ALMA) and the HIFI instrument \citep{Graauw2010} on board the Herschel Space Telescope \citep{Pilbratt2010}. 
Observations of the $^{13}$CO 3$\rightarrow$2 and C$^{18}$O 3$\rightarrow$2 lines were carried out using the Plateau de Bure Interferometer (PdBI) in its A configuration during October-November, 2014. This configuration provided a beam of 0.35$"$ $\times$ 0.18$"$ PA 25$^\circ$ ($\sim$63~au$\times$32~au at the Taurus distance). During the observations two 40 MHz correlator units were placed at the frequencies of the  $^{13}$CO  (330.588 GHz) and C$^{18}$O 3$\rightarrow$2 (329.330 GHz) lines providing a spectral resolution of 78 kHz. The C$^{18}$O 3$\rightarrow$2 line was not detected with an rms of 46 mJy/beam. Calibration and imaging were performed using the Grenoble Image and Line Data Analysis Software (GILDAS; \citealp{Pety2005}).  

Our study uses the  HCN 3$\rightarrow$2 line observations obtained from the ALMA archive. These observations
were performed during Cycle 4 within project number 2016.1.00813.S. 
The line was observed with a spectral resolution of $\sim$488.25 kHz and the synthesized beam is 0.07$''$ $\times$ 0.04$''$ with PA 27$^\circ$. This provides an 
unprecedented spatial resolution of $\sim$12~au$\times$7~au and the achieved  rms is 2.4 mJy/beam. We produced a 1.1~mm
continuum image with the channels and spectral windows empty of line emission. The continuum image has a synthesized beam of  0.06$"$ $\times$ 0.03$"$ with PA 27$^\circ$, and a rms of 0.13 mJy/beam.  The data were manually calibrated using the 4.7.1 version of the Common Astronomy Software Applications (CASA; \citealp{Mullin2007}) while the imaging process was performed using CASA version 5.4.1 following the National Radio Astronomy Observatory (NRAO) imaging guide. 

The Herschel Space Observatory \citep{Pilbratt2010} observations presented here
were obtained with the heterodyne instrument for the far infrared
HIFI \citep{Graauw2010} (OT1\_afuente\_2). The basic data reduction was performed using the standard
pipeline provided with the version 7.0 of HIPE \citep{Ott2010} and
then exported to GILDAS/CLASS \citep{Pety2005} for a more detailed
analysis.

\section{(Sub-)millimeter interferometric images} 
\label{sec:hyd}

Fig.~\ref{Fig1} shows the high angular resolution 1.1~mm continuum image obtained from ALMA data.  Based on a 2D Gaussian fit,  we
derive the flux, position and size of the compact continuum source.
The estimated flux is $\sim$48 mJy which is in good agreement with the previous measurement by \citet{Fuente2012}. 
The size of the compact source is 94$\pm$1 mas$\times$66$\pm$1 mas, which constrains the radius of
the dusty circumstellar disc  to R$\sim$8 au. In addition to the compact source, elongated emission is detected 
in the direction  of the outflow at $>$10$\sigma$  level (see Fig.~\ref{Fig1}).
It should be noted that the coordinates of the compact
continuum source are $\alpha$(J2000) $= 04^{\mathrm{h}}29^{\mathrm{m}}23^{\mathrm{s}}$.731, $\delta$(J2000) $=24^{\circ}33\arcmin01\arcsec$.11. This position is offset 0.19$''$  to the South as compared to the position measured by \citet{Fuente2012}, which would imply  proper motion of  $\sim$(0,$-$34) mas yr$^{-1}$.  The most recent estimation of the proper motions of GV Tau N are ($+$8.695$\pm$0.853) mas~yr$^{-1}$ in right ascension, and ($-$25.085$\pm$ 0.672) mas~yr$^{-1}$ in declination
\citep{GAIA2018}. Taking into account the lower angular resolution of  \citet{Fuente2012}, beam$\sim$0.5$''$, we consider that the observed displacement can be due to the proper motions of GV Tau N. The position of the continuum source is adopted as offset (0$''$,0$''$) through this paper.

\citet{Reipur2004} measured a 3.6 cm flux of 0.10$\pm$0.01 mJy towards GV Tau N. We estimate a flux
of 4.6$\pm$0.5 mJy for the extended component, which would imply a 3.6~cm to 1.1~mm spectral index, $\alpha^{3.6cm}_{1.1mm}$ $\sim +$1.1. 
This spectral index is higher than the one expected for an ionized spherical wind, $\alpha$$\sim$+0.6, and usually found in T Tauri and Herbig
Ae/Be stars \citep{Skinner1993, Alonso2009}. 
This higher spectral index is still consistent with an ionized wind but with unconventional assumptions, i.e a velocity power law 
 v(R)$\propto$R or a density law, n$_e$(R)$\propto$R$^{-3}$ \citep{Panagia1988}. 
We should also consider the possibility that radio emission has varied on time scales of tens of years. Variability is a common phenomenum 
in young T Tauri systems \citep{Ubach2017, Liu2017}.
It would be also possible that we have overestimated the flux coming from the jet because 
a fraction of the extended emission is coming from dust thermal emission. 
Simultaneous measurements of the flux at several wavelengths would be desirable to further constrain the cm-mm spectral index.  

Mid-infrared interferometry (VLTI/MIDI) observations to probe the circumstellar material around GV~Tau~N and S
were carried out by \citet{Roccatagliata2011}. They modeled the emission assuming the existence of two different black-body
components, which accounts for the different disc regions. They obtained that the emission within 1.5~au has a temperature of 
900~K, while a second colder emission at 150~K originates within 10~au from the central star. The sizes of these two components
are consistent with the size of the 1.1mm compact source as observed with ALMA for which we derive a radius R$\sim$8~au. 
Based on the measureded visibilities, they estimate an inclination, $i = 80^\circ$ for the GV~Tau~N disc. This value is consistent 
with GV~Tau~N being a highly obscured star.

Intense emission of the HCN~3$\rightarrow$2 line is detected towards GV~Tau~N. We have produced the zero-moment map using only the
channels with emission $>$3$\sigma$. Based on this map, we
estimate that the size of the HCN emission is R$\sim$0.18$"$, significantly larger than  the small compact continuum clump  (see Fig.~{\ref{Fig1}). 
The total velocity integrated flux within R$\sim$0.18$"$ is $\sim$1.4 Jy/beam$\times$km~s$^{-1}$. In order to check for
possible filtering in the HCN 3$\rightarrow$2 line map, we have compared ALMA fluxes with those measured by \citet{Fuente2012} using the PdBI.
\citet{Fuente2012} derived a total flux of 3.64$\pm$0.21 Jy$\times$km~s$^{-1}$ in a region of 0.32$''$$\times$ 0.25$''$. Therefore, the high spatial resolution ALMA observations recover $\sim$ 40\% of the flux detected in PdBI observations. This is consistent with the zero-moment map shown in Fig.~\ref{Fig1}, 
which presents evidence of a resolved component that extends towards the East.

\begin{figure*}
    \includegraphics [width=0.95\textwidth] {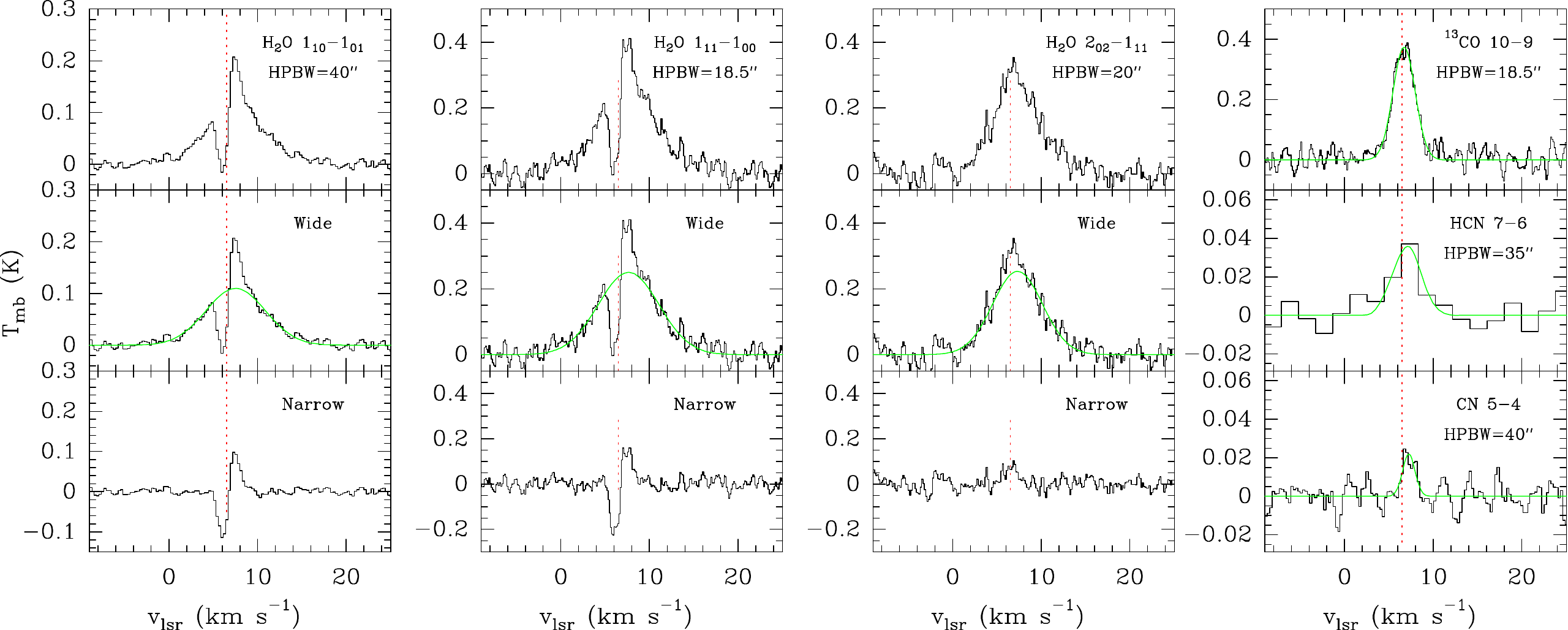}
    \caption{Spectra of the $^{13}$CO 10$\rightarrow$9, H$_2$O 1$_{11}$$\rightarrow$0$_{00}$, 1$_{10}$$\rightarrow$1$_{01}$ and 2$_{02}$$\rightarrow$1$_{11}$ lines obtained with the HIFI instrument on board the Herschel Space Observatory towards GV Tau N. The systemic velocity, v$_{\rm lsr}$=6.3~km s$^{-1}$, is indicated with the red dashed line.}
    \label{hifi}
\end{figure*}

\begin{figure}
    \includegraphics [width=0.45\textwidth] {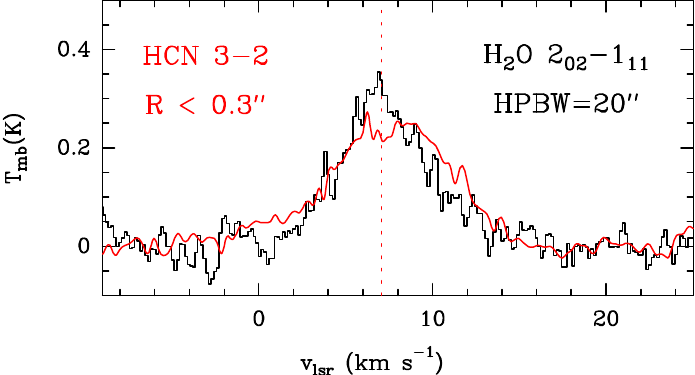}
    \caption{Spectra of the  H$_2$O 2$_{02}$$\rightarrow$1$_{11}$ line obtained with the HIFI instrument on board the Herschel Space Observatory towards GV Tau N. For comparison we have plotted the averaged ALMA spectrum of HCN 3$\rightarrow$2 in a circle of R$<$0.3$"$ around GV Tau N scaled to match the intensity of the H$_2$O 2$_{02}$$\rightarrow$1$_{11}$ line. The systemic velocity, v$_{\rm lsr}$=6.3~km s$^{-1}$, is indicated with the red dashed line.}
    \label{hifi2}
\end{figure}

An intense and asymmetric molecular ring is observed in the HCN 3$\rightarrow$2 image shown in Fig.~\ref{Fig1}.  
The inner and outer radii of this ring are R$_{\rm in}$$\sim$0.04$"$ (6 au)
and R$_{\rm out}$$\sim$0.18$"$ (17 au), respectively. In the azimuthal direction, the minimum of emission
is located in the outflow direction, suggesting that the jet has eroded the circumstellar material, sweeping out the
molecular gas. In Fig~\ref{Fig1} we also show the HCN 3$\rightarrow$2 spectra towards the star and the emission peak in the
molecular ring. The velocity profile towards the star is wide, with Full Width Half Power (FWHP) of $\sim$12~kms$^{-1}$, and asymmetric,
with the emission at highest velocities in the redshifted range. The spectrum towards the emission peak is
narrower with FWHP$\sim$10~km~s$^{-1}$, and presents two peaks located at blue-shifted and red-shifted velocities with respect to 
the systemic velocity, which is not expected in a pure rotation scenario. This wide line is not surprising and it is most likely due to the strong interaction between
the jet and the surrounding molecular gas. Also, self-absorption by the lower density envelope at the systemic velocity would produce the two-peak feature.

The  $^{13}$CO 3$\rightarrow$2 line is detected towards GV Tau N with a complex profile (see Fig.~\ref{Fig1}).
The emission presents high velocity wings at the same velocities as  the HCN 3$\rightarrow$2 spectrum. 
In addition, a narrow,  FWHP$\sim$2 km s$^{-1}$, absorption feature is observed at   v$_{\rm lsr}$ = 9.0 km s$^{-1}$, i.e.,  
red-shifted from the systemic velocity of v$_{\rm lsr}$ = 6.3 km s$^{-1}$ and consistent with the emission coming from infalling gas. It is interesting 
to note that \citet{Doppmann2008} detected HCN and C$_2$H$_2$ 
absorbing gas at  v$_{\rm lsr}$ = 8.7$\pm$0.3 km s$^{-1}$ (blue-dashed line in Fig.~\ref{Fig1}), which is close to the velocity of the
absorption detected in $^{13}$CO.  The signal-to-noise ratio  of our  $^{13}$CO 3$\rightarrow$2  observations does not allow to obtain a good quality image. 

\section{Far Infrared lines} 
\label{sec:FIR}

High excitation lines of $^{13}$CO, H$_2$O, HCN and CN  were observed with high spectral
resolution using the instrument HIFI onboard Herschel (see Fig.~\ref{hifi}). High signal-to-noise detections were obtained for the
$^{13}$CO 10$\rightarrow$9, and H$_2$O 1$_{11}$$\rightarrow$0$_{00}$, 1$_{10}$$\rightarrow$1$_{01}$ and 2$_{02}$$\rightarrow$1$_{11}$
lines. The HCN 7$\rightarrow$6 and CN 6$\rightarrow$5 lines were detected at 4$\sigma$ level  (see Fig.~\ref{hifi}). Unfortunately, the
Herschel angular resolution does not allow to resolve the two components of the binary, GV Tau N and GV Tau S, and we need
to use the kinematic information to discern the origin of the observed lines.

It is remarkable the difference between the profile of the $^{13}$CO 10$\rightarrow$9 line and those of the H$_2$O lines. 
The $^{13}$CO 10$\rightarrow$9 line presents a narrow profile, with FWHP$\sim$2.7~km~s$^{-1}$, and is
centered at v$_{\rm lsr}$=6.74$\pm$0.05 km s$^{-1}$. The  profiles of the ground ortho- and para-water lines are significantly wider,
and show a strong absorption at  v$_{\rm lsr}$$\sim$6.0~km~s$^{-1}$. 
The absorption is not observed in the higher excitation p-H$_2$O 2$_{02}$$\rightarrow$1$_{11}$ line (E$_u$/k$\sim$100 K).
Indeed, the profile of the water lines can be fitted using two components: i) a wide Gaussian with FWHP$\sim$ 8 ~km~s$^{-1}$, and (ii) a narrower
component  with FWHP$\sim$2.7~km~s$^{-1}$, similar to $^{13}$CO. Fig.~\ref{hifi} shows the decomposition of the water line profiles in
these two components that we hereafter name wide and narrow components. The Gaussian fits to these components are shown in Table~\ref{gaussians}.
It is specially interesting to discern whether the emission of the wide component arises in the compact gaseous disc detected with the ALMA maps or 
it is coming from a large scale outflow(s). 
In order to explore the origin of this component, we compare its profile with  the mean HCN 3$\rightarrow$2 spectrum in the region R$<$0.3$"$ (see Fig.~\ref{hifi2}).
The wings of the p-H$_2$O 2$_{02}$$\rightarrow$1$_{11}$ line are similar to those of the interferometric HCN 3$\rightarrow$2 line, suggesting that
the origin might be similar.  We further discuss the origin of this component in Section~6.

In the bottom panels of Fig.~\ref{hifi} we show the profiles of the water narrow component. This component presents absorption at blue-shifted velocities. 
We recall that the absorptions observed in the NIR observations lie at red-shifted velocities, $\sim$8.7 km~s$^{-1}$.
An absorption at similar velocity, $\sim$9~km~s$^{-1}$, has been observed in our PdBI spectrum of the $^{13}$CO 3$\rightarrow$2 line. 
Thus, the blue absorption features detected in the FIR water lines are  more likely caused by a colder layer in the outer envelope.
Fig.~\ref{hifi} also shows the spectra of the HCN 7$\rightarrow$6 and CN 5$\rightarrow$4 lines.
These spectra only present the narrow component. We cannot discard, however, the existence of a weaker wide component that it is not detected because of
the low S/N of the detections. This should be the case of HCN 7$\rightarrow$6, for which emission presents a slightly wider spectrum. It should be noticed that
the peak of the weak and narrow HCN 7$\rightarrow$6 lines is at the velocity of the dip observed in the ALMA HCN 3$\rightarrow$2 profile, suggesting that the narrow line
observed with Herschel is probing the lower density envelope.

\begin{figure*}
    \includegraphics [width=0.97\textwidth] {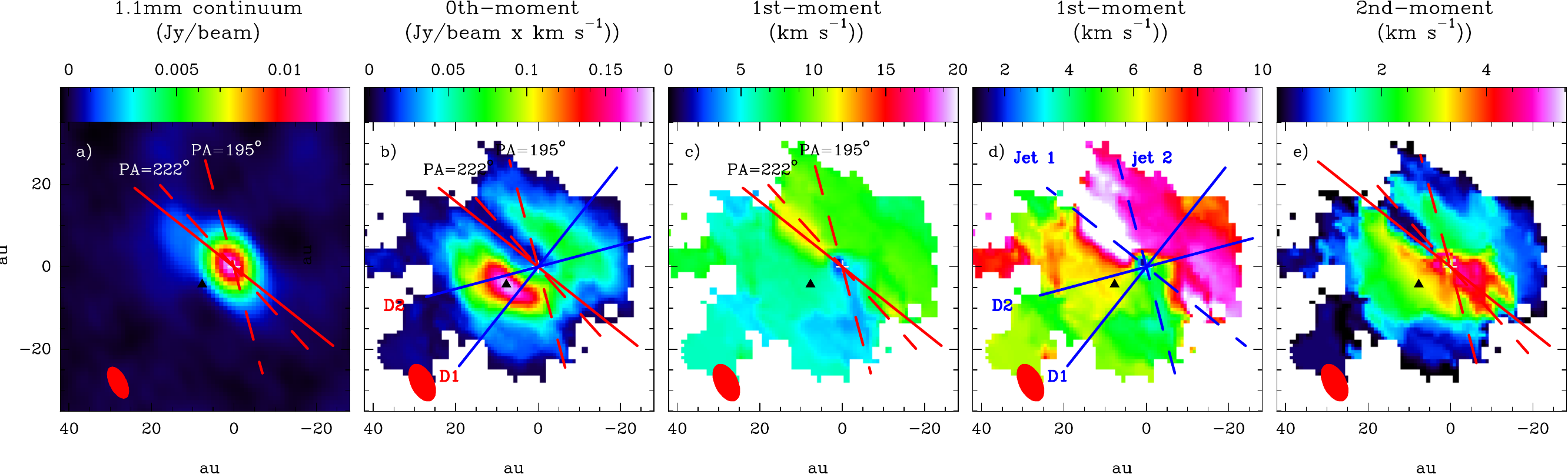}
    \caption{Continuum image at 1.1 mm and moment maps of the HCN 3$\rightarrow$2 emission as observed with ALMA. Dashed lines indicate  the 
    direction of the giant Herbig Haro flow at PA = 222$^\circ$ \citep{Devine1999} and of the optical jet detected at PA=195$^\circ$ \citep{Movsessian1999}. 
    We also indicate the directions perpendicular to $Jet~1$, named $D1$, and to $Jet~2$, named $D2$.
    Panels {\it c)}  and {\it d)} show the same map with different colour scales
    to highlight the bipolar outflow(s) ({\it c)}) or the rotation motion ({\it d)}). 
    The directions  $Jet~1$, $Jet~2$, $D1$ and $D2$ are selected to perform the position-velocity  diagrams.}
    \label{Fig3}
\end{figure*}

\section{Kinematics of the molecular gas} 
\label{sec:kin}

In Fig.~\ref{Fig3} we present the first and second moment maps of the HCN 3$\rightarrow$2 emisison as observed with ALMA. 
The emission at high red- and blue-shifted velocities, $|v-v_0| >$ 4 km s$^{-1}$, shows the presence of two misaligned outflow lobes:
the red lobe is close to the direction defined by the giant Herbig Haro flow,  PA=222$^\circ$ \citep{Devine1999} (hereafter, $Jet~1$), and
the blue lobe is placed at PA=195$^\circ$, the direction of the Herbig Haro jet detected by \citet{Movsessian1999} (hereafter, $Jet~2$). 
This misalignment can be interpreted as the consequence of a precessing bipolar outflow. 
Alternatively, these lobes might belong to two different bipolar outflows.

In order to discern between these two scenarios we inspect the channel maps shown  in Fig.~\ref{Vel1} and Fig.~\ref{Vel2}. The directions of $Jet~1$ and
$Jet~2$ are indicated in each panel.  The red lobe of $Jet~1$ is clearly seen in  panels  $v_{lsr}$=10.28~km~s$^{-1}$ to 13.36~km~s$^{-1}$,
located SW from the star. The blue counterpart is missing in the HCN 3$\rightarrow$2 channels maps. The molecular outflow
is formed by the circumstellar matter that is being swept up by the ionized jet. The lack of a blue lobe is due to the 
absence of the molecular gas in this direction, where a wide cavity has already being excavated by the jet, and the molecular gas has been cleaned. 
The existence of high velocity gas along $Jet~2$ is not such clear since the kinematic signatures of the blue and red lobes are mixed with the 
disc rotation.  High velocity gas along $Jet~2$ is detected  in panels  $v_{lsr}$=$-$1.48~km~s$^{-1}$ to 2.16~km~s$^{-1}$.  
 Fig.~\ref{Vel2} shows that high velocity gas is detected in panels $v_{lsr}$=$-$13.08~km~s$^{-1}$ to 14.48.16~km~s$^{-1}$ 
 which is located in the $Jet~2$ direction and can be interpreted as the red lobe of this outflow. Given the angular resolution of the observations,
 it might also be associated to the walls of the cavity excavated by $Jet 1$.

Fig.~\ref{FigPV} shows the position-velocity (p-v) diagrams along  $Jet~1$ and $Jet~2$. 
The $Jet~1$ p-v diagram  presents clumpy emission with small changes in the velocity between one small bright clump and the following. 
These bright spots are also observed as  narrow features in the HCN 3$\rightarrow$2 spectrum shown in Fig~\ref{Fig1}.
This kind of "saw-like" features have been detected in high velocity jets and were interpreted as
oblique shocks formed in the interphase between the high velocity jet and the circumstellar matter \citep{Santiago2009}. In this scenario, the 
clumps at red-shifted and blue-shifted velocities correspond to the back and front walls 
of the cavity excavated by the jet in the gaseous disc.  This interpretation is also consistent with the morphology 
observed in the zero moment map of  the HCN 3$\rightarrow$2 emission (see Fig.~\ref{Fig1}).
The outflow $Jet~1$ crosses the  molecular ring through the azimuthal  minimum of the HCN 3$\rightarrow$2 emission, probing
that the  gas has been swept up.

The p-v diagram along $Jet~2$ is also shown in Fig.~\ref{FigPV}.The outflow $Jet~2$ presents bipolar morphology with the highest 
velocities close to the star. In this case, the blue lobe is more intense than the red lobe, contrary to the case of $Jet~1$.  
However, an intense clump is detected at 10~km~s$^{-1}$ in the direction of $Jet~2$ (R1 in Fig.~\ref{FigPV}) that might be disc gas accelerated by
the $Jet2$. However, we need to be cautious in our conclusions because the sensitivity and angular resolution of the HCN 3$\rightarrow$ observations
do not allow to fully disentangle $Jet~2$ from the gas that is accelerated in the walls of the cavity excavated by $Jet~1$.

\begin{figure*}
    \includegraphics [width=0.97\textwidth] {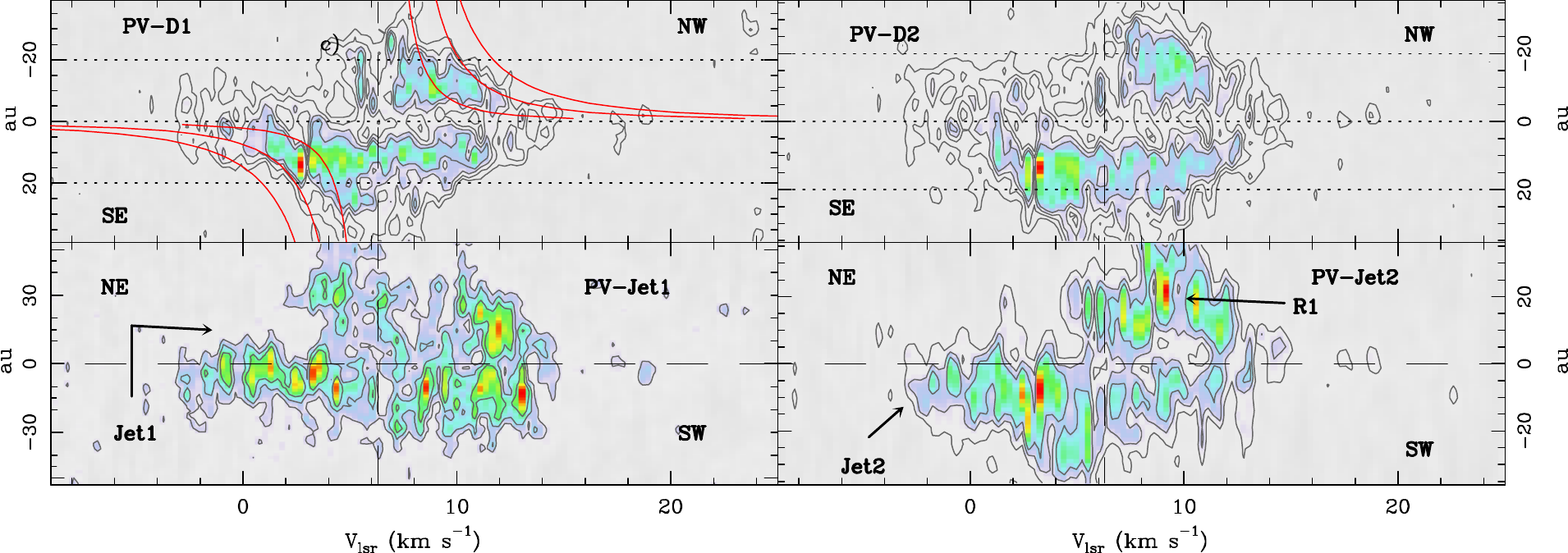}
    \caption{ Position-velocity diagrams of the HCN 3$\rightarrow$2 line along the strips  $Jet~1$, $Jet~2$, $D1$ and $D2$. In panel $D1$, we show the Keplerian velocity curves assuming M$_*$=0.8 M$_\odot$ and  i=20$^\circ$, 40$^\circ$, and 65$^\circ$ (i=0$^{\circ}$ would be face-on) for the circumbinary disc. Observational data fits better the i=40$^\circ$ curve. Contour levels are 2$\sigma$ to 5$\sigma$ in steps of 1$\sigma$ ($\sigma$=2.4~mJy/beam).}
    \label{FigPV}
\end{figure*}

In order to explore the kinematics of the circumbinary disc, we performed the p-v diagrams along the directions perpendicular to
$Jet~1$ and $Jet~2$, named $D1$ and $D2$, respectively (Fig.~\ref{FigPV}). The p-v diagrams along $D1$ and $D2$ present emission in the four quadrants, which
is not compatible with a pure rotating disc. A combination of rotation and outflow is necessary to explain 
this pattern. Furthermore, there is a gap around the position of the stars with a size, R$\sim$8~au, in both diagrams. 
In spite of the highly disturbed kinematics, we can see a hint for Keplerian rotation in the p-v diagram along $D1$. 
For comparison we compare the observations with the Keplerian curves for a star with M$_*$=0.8 M$_\odot$ and  
inclination angles i=20$^\circ$, 40$^\circ$, and 65$^\circ$. We find reasonable agreement between the observations in the inner R$<$20~au region
and the Keplerian curves assuming  i=40$^\circ$ curve. This supports the interpretation that the HCN 3$\rightarrow$2 line comes from 
a rotating disc which kinematics is highly perturbed by the interaction with the bipolar outflow(s). 
We recall that there is a degeneracy between the stellar mass and the inclination angle.
If GV~Tau~N is a binary, the stellar mass content would be larger, and a lower inclination angle would be required to fit the observations.
This would be consistent with the observed morphology, which is that expected for an almost face-on disc. 

\citet{Doppmann2008} proposed that GV Tau N could be a binary  to explain the large difference between the velocity of the lines coming from the stellar 
photosphere and the molecular gas. The existence of two bipolar outflows would also explain the kinematics observed in the ALMA 
HCN 3$\rightarrow$2 cube. The most energetic outflow, $Jet~1$, should be in (or close to) the equatorial plane of the circumbinary disc in order to explain 
the cavity cleaned of gas observed in the HCN maps. The second outflow would be perpendicular, or close to the perpendicular direction, to the circumbinary disc, hence flowing away with little interaction with the molecular gas (see a sketch in Fig.~\ref{Sketch}). This geometry would imply that the two components of the binary are highly misaligned.  

Binarity could also explain the different morphology of the gas and dust around GV Tau N, the size of the dust continuum emission being smaller. Stellar binaries exert gravitational torques on inclined circumbinary discs causing radially differential precession and, under some circumstances, warps
in the disc. Large dust grains in thick discs pile up at the warp location, forming narrow dust rings within an extended gaseous disc  \citep{Aly2020}.
This warped narrow dust ring could be the emitting region of the T=150~K component detected in the mid-IR interferometric observations by
\citet{Roccatagliata2011}.

Misaligned binaries are common in young stellar systems (see e.g. \citealp{Brinch2016, Takakuwa2017}).
GV Tau S was also proposed to be a spectroscopic binary by  \citet{Doppmann2008}. If confirmed, GV Tau would be a quadruple star system such as
GG Tau \citep{Andrews2014} and HD98800B \citep{Kennedy2019}.

\section{Molecular chemistry} 
\label{sec:cd}

In this Section we compile the column densities and rotation temperatures that have been derived 
using observations from NIR to mm wavelengths (see Table~\ref{cd}). Even when comparing
lines of the same species,  ro-vibrational and rotational lines are emitted from different regions of this complex
system and the comparison is not straightforward. Our goal is to put together  all the pieces of information
to figure out the overall scenario.

The physical and chemical conditions in the inner R$<$1$-$10~au are better probed by NIR observations.
Different rotation temperatures and molecular column densities have been measured with NIR molecular absorption 
observations towards GV Tau N (see Table~\ref{cd}). 
In particular, \citet{Gibb2007} derived T$_{rot}$= 115$\pm$11~K   and  N(HCN)= (3.7 $\pm$ 0.3) $\times$ 10$^{16}$ cm$^{-2}$
in the first detection of the HCN absorption lines towards this star. Later, \citet{Doppmann2008} and \citet{Bast2013} estimated
a higher rotation temperature, T$_{\rm rot}$$>$500~K for HCN (see Table~\ref{cd}). 
\citet{Doppmann2008} argued that this difference is due to the different set of lines used in
the calculations. \citet{Gibb2007} calculations were based on lines at lower energy and obtained a lower rotation temperature, 
as expected in the case of a steep gas kinetic temperature gradient along the line of sight.
The velocity  centroid of the C$_2$H$_2$ and HCN absorption lines is v$_{\rm lsr}$ = 8.7 $\pm$0.3 km s$^{-1}$, similar to
that of the absorption feature that we have detected in the $^{13}$CO 3$\rightarrow$2 line on the top of high velocity 
wings, thus consistent with the scenario of these molecules coming from infalling gas towards the star. 
 \citet{Gibb2007, Gibb2008} also derived  $^{12}$CO, $^{13}$CO, and C$^{18}$O  rotation temperatures based on molecular absorption (see Table~\ref{cd}).
They obtained  T$_{rot}$= 140~K $-$ 260~K,  suggesting that the observed lines are probing molecular gas located further from the star than HCN and C$_2$H$_2$
although along the line of sight.

\begin{figure}
\includegraphics [width=0.5\textwidth] {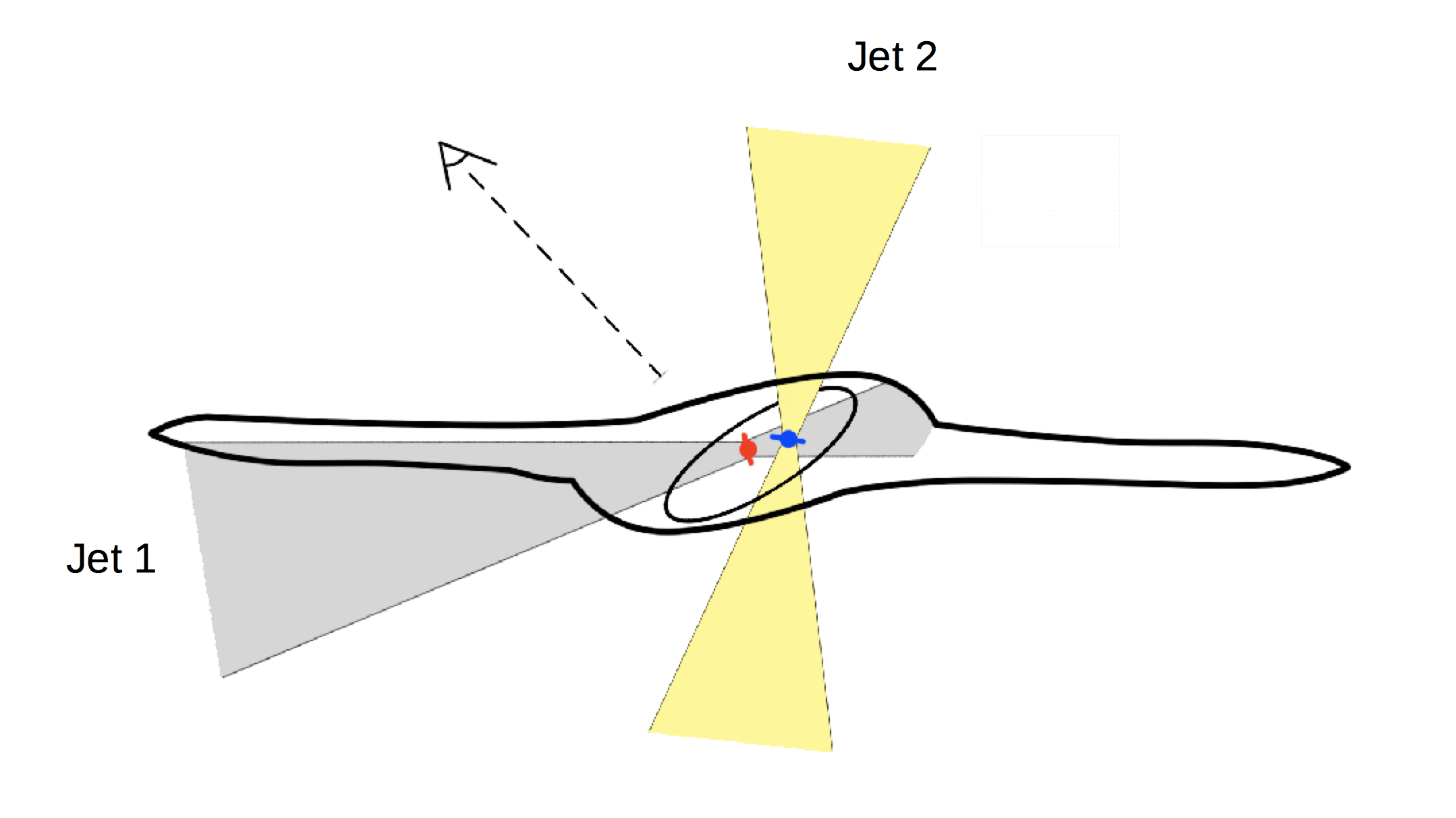}
    \caption{Sketch of the geometry proposed for GV Tau N to explain ALMA spectroscopic observations.}
    \label{Sketch}
\end{figure}

We present the first detection of water towards GV Tau (see Fig.~\ref{hifi}). In particular, three water lines, p-H$_2$O 1$_{11}$$\rightarrow$0$_{00}$, 
o-H$_2$O 1$_{10}$$\rightarrow$1$_{01}$ and p-H$_2$O 2$_{02}$$\rightarrow$1$_{11}$ have been detected. 
The emission of the ground ortho and para water lines are thought to originate in the disc surfaces and in shocks along the outflow 
cavity in Class I sources  \citep{Hogerheijde2011, Kristensen2012, Karska2013, Podio2013, Mottram2014, Salinas2016}.
The FIR water lines show two well differentiated components: one with a wide profile, very similar to that derived 
from the interferometric HCN 3$\rightarrow$2 observations; the second one is characterized by a narrow profile and was interpreted as coming from the outer envelope.  
In the following, we explore the possibility that the emission of the wide component comes from the GV Tau N disrupted gaseous disc 
detected with  ALMA observations of the HCN 3$\rightarrow$2 line.  
First, we would assume that the bulk of the water emission is coming from GV Tau N. This is a reasonable assumption taking into account
that previous NIR and millimeter observations showed that molecular absorption/emission is mainly coming from the
North component \citep{Gibb2007, Gibb2008, Doppmann2008, Bast2013, Fuente2012}. Secondly, we use the two observed p-H$_2$O line to 
investigate the physical conditions of the emitting region. Based on our Gaussian fits to the wide component (Table{\ref{cd}), and after correcting
for the small difference in the beam size, we  derive an integrated intensity ratio,
 R$_{\rm obs}$(H$_2$O@987~GHz/1113~GHz)= 1.0$\pm$0.1 for the two p-H$_2$O lines observed. This ratio is not consistent
with optically thin emission. In the optically thin limit, and assuming Local Thermodynamic 
Equilibrium (LTE), we estimate R$_{\rm LTE}$(H$_2$O@987~GHz/1113~GHz)=0.5 at 200~K, and increases with temperature until 
$\sim$0.63 for  T$_k$$>$500~K.  The observed ratio can only be explained if the emission of p-H$_2$O is optically thick. This would imply 
N(p-H$_2$O)$>$ a few 10$^{17}$~cm$^{-2}$. This high value of the water column density is within the 
range of water column densities derived towards T Tauri stars using Spitzer observations  \citep{Salyk2007}. Thus,
it is not unreasonable that the emission comes from the compact disrupted gaseous disc. To further check the consistency of our interpretation, we perform
the following calculation: assuming T$_{\rm rot}$$\sim$ 500~K as derived from HCN absorption 
observations, optically thick emission, and a source size of 0.5$"$ (see Fig.~\ref{Fig1}),
we  predict T$_{\rm mb }$$\sim$ 0.23~K for the p-H$_2$O lines after diluting to the Herschel beam, which is fully consistent with Herschel observations.  

It is also interesting to compare the intensities of the ground  p-H$_2$O line with the o-H$_2$O 1$_{10}$$\rightarrow$1$_{01}$ line at 557~GHz. 
In the optically thick case, we would expect R$_{\rm thick}$(H$_2$O@1113~GHz/557~GHz)$\sim$4.7 because of the different observational beams. Instead, the observed value is  R$_{obs}$(H$_2$O@1113~GHz/557~GHz)$\sim$2.4$\pm$0.1. This can be explained if the size of the emitting
region is larger for  ortho-H$_2$O than for the para-H$_2$O.  Indeed, this is expected according to the critical densities of these line 
($\sim$2$\times$10$^7$ cm$^{-3}$ and $\sim$2$\times$10$^8$ cm$^{-3}$ at 50 K for the 557 and 1113 GHz lines, respectively),  and
the higher abundance of o-H$_2$O (the ortho-to-para ratio is $\sim$3 at high temperatures). Direct imaging of these lines at 
high angular resolution is required to confirm this scenario. 

We have also derived $^{13}$CO, HCN and CN column densities based on FIR observations. In these cases, the lines show narrow 
profile and therefore, we assume a lower rotation temperature T$_{\rm rot}$=250~K, similar to those measured from NIR absorption 
lines of $^{13}$CO \citep{Gibb2007, Gibb2008}. Again, the unknown source size introduces a large uncertainty in our calculations. 
In order to make a fair comparison between molecules, we assumed a source size of 18.5$"$.
With these assumptions we derive the following source-averaged column densities: N($^{13}$CO)=5.0$\times$10$^{14}$~cm$^{-2}$,
N(HCN)= 4.0$\times$10$^{11}$~cm$^{-2}$, and N(HCN)= 6.0$\times$10$^{11}$~cm$^{-2}$.

\begin{table}
\caption{Summary of molecular observations}
\label{cd}
\begin{tabular}{lccc} \hline 
 \multicolumn{4}{c}{NIR molecular absorption} \\
 \multicolumn{1}{c}{Species}  &  \multicolumn{1}{c}{T$_{\rm rot}$(K)}  &  \multicolumn{1}{c}{N(cm$^{-2}$)}   & \multicolumn{1}{c}{Ref}   \\  \hline   \hline 
$^{12}$CO        &   200$\pm$40  &  $\sim$1.2$\times$10$^{19}$                & (1) \\
$^{13}$CO        &   260$\pm$20  &  $\sim$1.1$\times$10$^{17}$                & (1) \\
C$^{18}$O       &   140$\pm$50  &  (1.4 $\pm$ 0.5) $\times$ 10$^{16}$     & (1)  \\
C$_2$H$_2$    &   170$\pm$20  &   (7.3 $\pm$ 0.2) $\times$ 10$^{16}$     & (1) \\
                         &   720$\pm$60  &   (1.4 $\pm$ 0.3) $\times$ 10$^{16}$     &  (3) \\
CH$_4$            &   750$\pm$50  &   (2.8$\pm$0.2)$\times$10$^{17}$         &  (4) \\
CO$_2$           &    250$\pm$25  &   (5.1 $\pm$ 0.7) $\times$ 10$^{16}$     & (3) \\  
HCN                 &   115$\pm$11    &     (3.7 $\pm$ 0.3) $\times$ 10$^{16}$   & (1)  \\ 
                        &   440$\pm$40   &     (1.8 $\pm$ 0.4) $\times$ 10$^{16}$   & (3) \\  
                        &    550                &    1.5 $\times$ 10$^{17}$                         & (2)  \\
                         \hline
 \multicolumn{4}{c}{FIR molecular emission} \\
 \multicolumn{1}{c}{Species}  &  \multicolumn{1}{c}{T$_{\rm rot}$(K)}  &  \multicolumn{1}{c}{N(cm$^{-2}$)}   & \multicolumn{1}{c}{Comments}   \\  \hline     
 o-H$_2$O   &  500   &   $>$10$^{17}$  &  Wide   \\
 p-H$_2$O   &  500   &   $>$10$^{17}$  &  Wide  \\
$^{13}$CO   &  250   &   5.0$\times$10$^{14}$  &  Narrow $^1$  \\
HCN            &  250    &   4.0$\times$10$^{11}$  &  Narrow$^1$   \\
CN              &  250    &   6.0$\times$10$^{11}$  &  Narrow$^1$   \\   \hline    
 \multicolumn{4}{c}{mm interferometric emission} \\
 \multicolumn{1}{c}{Species}  &  \multicolumn{1}{c}{T$_{\rm rot}$(K)}  &  \multicolumn{1}{c}{N(cm$^{-2}$)}   & \multicolumn{1}{c}{Comments}   \\  \hline     
$^{13}$CO   &  250$-$500   &   (2.6$-$5.0)$\times$10$^{17}$  & Wide  \\
HCN            &  250$-$500    &   (2.0$-$3.5)$\times$10$^{15}$  &  Wide   \\   \hline \hline
\end{tabular}

\noindent
References: (1) \citealp{Gibb2007, Gibb2008}; (2) \citealp{Doppmann2008}; (3) \citealp{Bast2013}; (4) \citealp{Gibb2013}\\
\noindent
$^1$ Beam averaged column densities  (beam=18.5$"$).
\end{table}

Finally, we explore the chemistry of the compact emission around GV~Tau~N using the millimeter interferometric images.
\citet{Fuente2012} concluded that the HCN/HCO$^+$ abundance ratio should be $>$300 to be compatible with the non-detection
of HCO$^+$ 3$\rightarrow$2 with PdBI observations. Here, we will compare the HCN 3$\rightarrow$2 observations obtained with
ALMA with the PdBI observations of the $^{13}$CO $\rightarrow$2 line.  For this comparison we use the averaged HCN 3$\rightarrow$2
spectrum in R$<$0.18$"$, since the beam of $^{13}$CO 3$\rightarrow$2 observations is $\sim$ 0.3$"$.  We can fit the PdBI 
$^{13}$CO 3$\rightarrow$2 line profile, assuming LTE, T$_{\rm rot}$=250$-$500~K and N($^{13}$CO)=(2.6$-$5.0)$\times$10$^{17}$ cm$^{-2}$. At these
high temperatures, the emission of the $^{13}$CO 3$\rightarrow$2 is optically thin and it is not very dependent on the assumed gas
kinetic temperature. This estimate is in reasonable agreement with the value derived from the NIR absorption lines, suggesting that we are 
probing the same gas component. The R$<$0.18$"$ averaged profile of the HCN 3$\rightarrow$2 emission can be fitted
with  T$_{\rm rot}$=250$-$500~K and N(HCN)=(2.0-3.5)$\times$10$^{15}$ cm$^{-2}$, i.e., two orders of magnitude lower than the values
derived from the NIR molecular absorptions. This suggest that the mm emission is probing a different gas component than
the molecular absorption in NIR. Molecular absorptions arise in a very small and hot region where
the HCN lines are optically thick, and T$_{\rm b}$ (HCN 3$\rightarrow$2) $\sim$ 500 K. In order to be compatible with ALMA observations,
the size of this region should be R$\sim$ 2$-$3 au around the star, confirming its origin in the GV Tau N individual disc.

The observation of $^{13}$CO and HCN with different techniques provides valuable information on the chemical 
changes occurring in this complex system. As shown in Table~\ref{cd}, we obtain that the HCN/$^{13}$CO abundance ratio is $\sim$10$^{-3}$ based on
the FIR observations, $\sim$10$^{-2}$ in the gaseous disc traced with ALMA and PdBI observations, and $\sim$1 in the close vicinity
of the star where NIR molecular absorptions arise. These different values of the  HCN/$^{13}$CO ratio testify for chemical gradients
in this system, with the HCN/$^{13}$CO ratio decreasing with the distance from the star.
This behavior of the HCN abundance, peaking at R$\sim$2-3 au from
the star is consistent with chemical models predictions (see e.g. \citet{Walsh2010, Agundez2018}). These models
also predict that a high abundance of H$_2$O is expected in this inner region, coeval with HCN.  The narrow component appearing
in the H$_2$O line profiles is thought to come from a more extended envelope component. 

\section{Concluding remarks} 

The high spatial resolution of ALMA observations allows us, for the first time, to discern 
the different gas components in the vicinity of GV Tau N.
The young star GV Tau N is located in the inner gap of a gaseous Keplerian disc of R$\sim$20~au. An ionized jet is detected in the continuum at 1.1~mm. 
Furthermore, ALMA observations present evidence of the existence of two molecular outflows, one of them associated with the continuum jet.
We propose that all these observations can be easily explained if GV Tau N is a binary, as already 
suggested by \citet{Doppmann2008}. The individual disc of the primary is highly misaligned relative to the circumbinary disc, and its jet
is disrupting the molecular ring (see Fig.~\ref{Sketch}).

The present detection of the o-H$_2$O and p-H$_2$O line in GV Tau N is of paramount importance.
Determining the water abundance in proto-planetary discs has been remarkably difficult.
Water vapor detections using the ground state H$_2$O lines have been reported for just two discs,
the well-known transition disc TW Hya \citep{Hogerheijde2011,Salinas2016},  and the Class I disc DG Tau \citep{Podio2013}.
Stringent upper limits to the water abundance have been reported for a dozen other sources at an order of magnitude lower than
expected \citep{Bergin2010, Du2017}. Recently, \citet{Harsono2020} unsuccessfully searched for water in five Class I discs 
using ALMA observations of millimeter lines of the rarer isotopologue H$_2^{18}$O. The detection of cold water towards
Class I protostar GV Tau stands as a  major result that deserves further study. 

We also present observations of high excitation lines of $^{13}$CO, HCN and CN as observed
with the HIFI/Herschel instrument, which provide information on the column densities of these
species at spatial scales of $\sim$ a few 10$"$.  As shown in Table~\ref{cd}, the HCN/$^{13}$CO abundance ratio is $\sim$10$^{-3}$ in the 
warm envelope,
$\sim$10$^{-2}$ in the gaseous disc better traced with ALMA and PdBI observations, and $\sim$1 in the close vicinity
of the star where NIR molecular absorptions arise. This behavior of the HCN abundance, peaking at R$\sim$2-3 au from
the star is consistent with chemical models predictions (see e.g. \citet{Walsh2010, Agundez2018}). These models
also predict that a high abundance of H$_2$O is expected in this inner region, coeval with HCN. 

Spitzer observations revealed that GV~Tau~N is specially rich in organic species \citep{Gibb2007, Gibb2008, Doppmann2008, Bast2013}.
The interaction of the jet with the circumbinary gaseous disc is expected to produce intense shocks that 
could contribute to enhance the abundance of organic species in this disc. High spatial resolution interferometric observations are required to
disentangle the interplay between chemistry and dynamics in this prototypical system.



\section*{Data availability}
The data underlying this article will be shared on reasonable request to the corresponding author.

\section*{Acknowledgements}
This paper makes use of the following ALMA data: ADS/JAO. 2016.1.00813.S.
ALMA is a partnership of ESO (representing its member states), NSF (USA) and NINS (Japan), together with NRC (Canada), 
MOST and ASIAA (Taiwan), and KASI (Republic of Korea), in cooperation with the Republic of Chile. The Joint ALMA Observatory 
is operated by ESO, AUI/NRAO and NAOJ. We thank the Spanish MINECO for funding support from AYA2016-75066-C2-2-P. SPTM has received funding  
from the European Union's Horizon 2020 research and innovation program under grant agreement No 639459 (PROMISE).







\appendix

\section{Supporting material}

\begin{table*}
\caption{Gaussian fits to FIR observations}
\label{gaussians}
\begin{tabular}{lrccccc} \hline 
 \multicolumn{1}{c}{line}  &  \multicolumn{1}{c}{Freq}  &  \multicolumn{1}{c}{beam}   & 
 \multicolumn{1}{c}{Area} &   \multicolumn{1}{c}{V$_{\rm lsr}$} &  \multicolumn{1}{c}{FWHP} &   \multicolumn{1}{c}{T$_{mb}$} \\  
  \multicolumn{1}{c}{}  &  \multicolumn{1}{c}{(GHz)}  &  \multicolumn{1}{c}{(")}   & 
 \multicolumn{1}{c}{(K km~s$^{-1}$)} &   \multicolumn{1}{c}{(km~s$^{-1}$)} &  \multicolumn{1}{c}{(km~s$^{-1}$)} &   \multicolumn{1}{c}{(K)} \\  
 \hline \hline    
o-H$_2$O   1$_{10}$$\rightarrow$1$_{01}$   &  556.94      &  40      &   0.93 ( 0.02)   &    7.52 ( 0.09)  &   7.92 ( 0.23)   &   0.11  \\
p-H$_2$O   1$_{11}$$\rightarrow$0$_{00}$   & 1113.34  &  18.5   &   2.14 ( 0.05)   &   7.58 ( 0.08)  &    7.42 (0.20)   &    0.27  \\
p-H$_2$O  2$_{02}$$\rightarrow$1$_{11}$    &  987.93       &   20     &  1.79  ( 0.06)   &   7.32 ( 0.09)  &    6.64 (0.29)  &   0.25  \\
$^{13}$CO 10$\rightarrow$9                           & 1101.35         &  18.5   &  1.18  ( 0.03)   &   6.72 ( 0.03)  &   2.97 ( 0.08)   &  0.37  \\
 CN             5$\rightarrow$4                              &  566.95       &  40      &   0.04 ( 0.01)   &    7.25 ( 0.13)  &   1.65 ( 0.25)  &  0.02 \\
HCN            7$\rightarrow$6                             &  620.30      &  35     &   0.14 ( 0.03)   &   7.02 ( 0.37)  &   3.71 ( 1.14)  &  0.04 \\
   \hline \hline
\end{tabular}
\end{table*}

\newpage

\begin{figure*}
\includegraphics [width=0.95\textwidth] {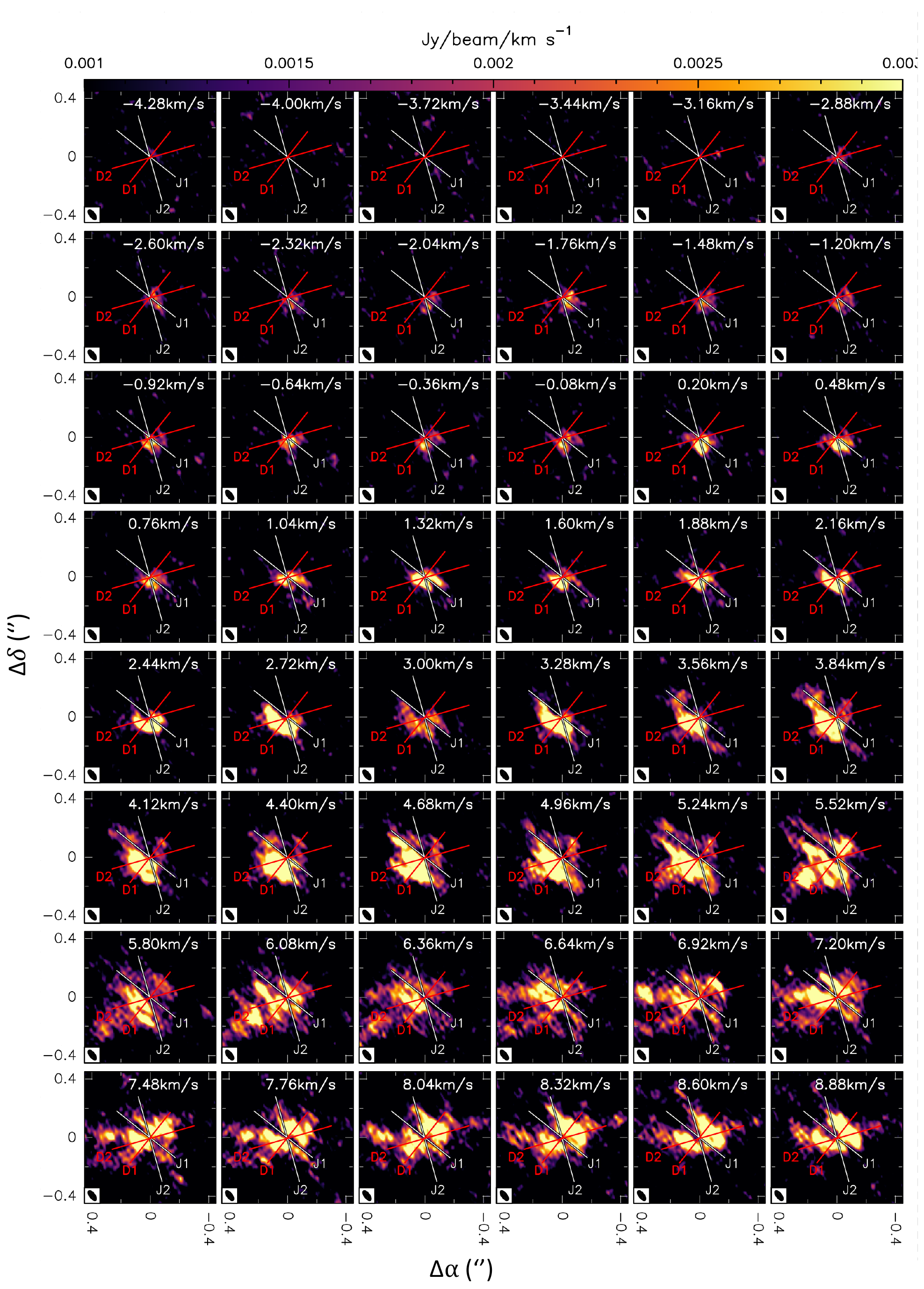}
    \caption{Channel maps of the HCN 3$\rightarrow$2 line as observed with ALMA. Solid lines indicate the directions of $Jet~1$, $Jet~2$, $D1$ and $D2$. }
    \label{Vel1}
\end{figure*}

\begin{figure*}
\includegraphics [width=0.95\textwidth] {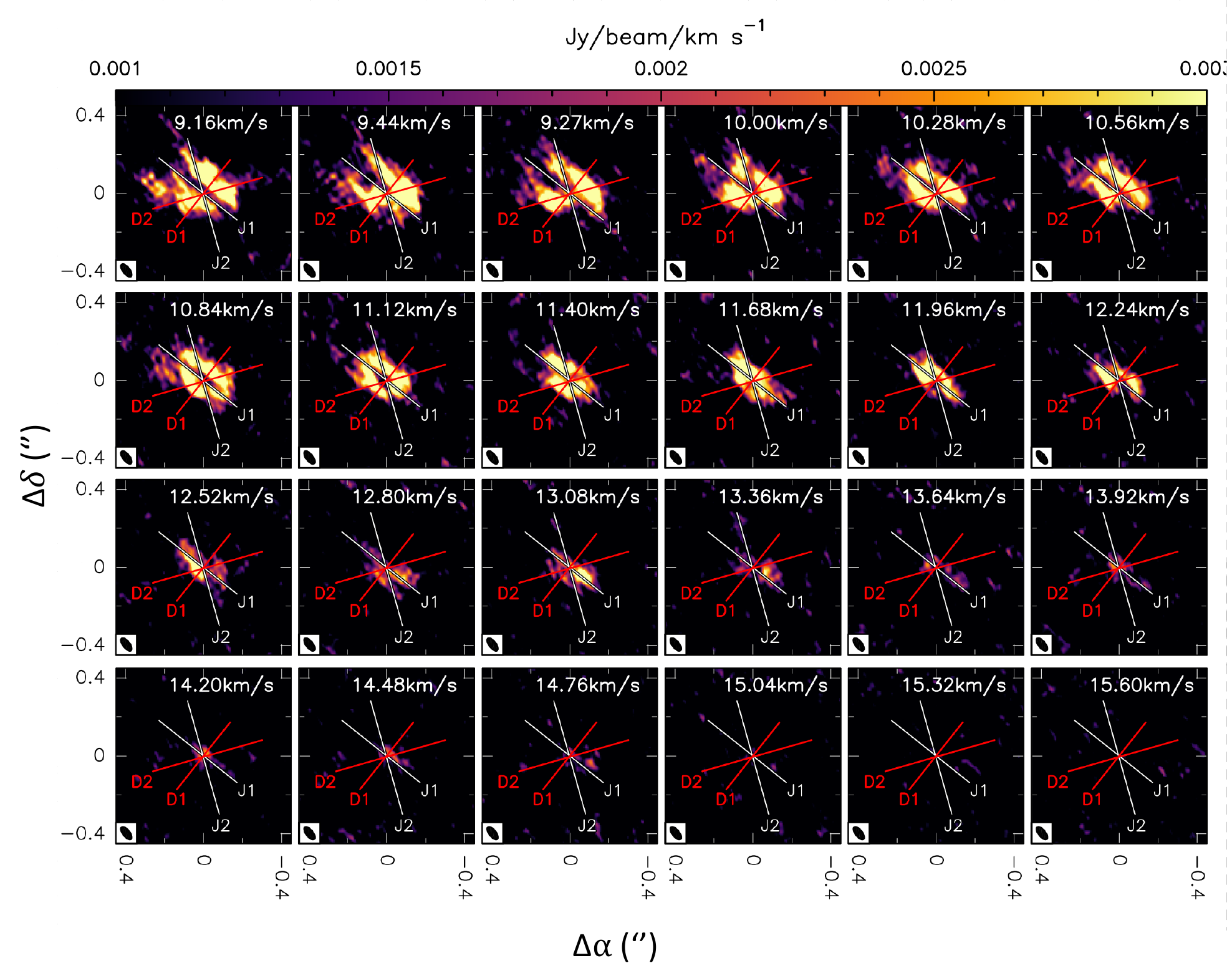}
    \caption{The same as Fig.~\ref{Vel1}.}
    \label{Vel2}
\end{figure*}



\bsp	
\label{lastpage}
\end{document}